\documentclass[prl,twocolumn,showpacs]{revtex4}
\usepackage{epsfig}
\def\avg#1{\langle#1\rangle} 
\def\Re{\rm{Re}}
\def\Im{\rm{Im}}
\def\be{\begin{equation}}       \def\ee{\end{equation}}
\def\bea{\begin{eqnarray}}      \def\eea{\end{eqnarray}}

\def\PRA{Phys. Rev. A}
\def\PRB{Phys. Rev. B}

\def\PRL{Phys. Rev. Lett.}

\begin{document}
\title{Competing orders in one dimensional spin 3/2 fermionic systems}
\author{Congjun Wu}
\affiliation{Department of Physics, Stanford University, Stanford,
CA 94305}
\affiliation{Kavli Institute for Theoretical Physics, University
of California, Santa  Barbara, CA 93106}
\begin{abstract}
Novel competing orders are found in spin 3/2 cold atomic
systems  in 
one-dimensional optical traps and lattices.
In particular, the quartetting phase,  a four-fermion counterpart of
Cooper pairing, exists in a large portion of the phase diagram.
The transition between the quartetting and singlet Cooper 
pairing phases is controlled by an Ising symmetry breaking effect
in one of the spin channels.
The singlet Cooper pairing phase also survives in the
purely repulsive interaction regime.
In addition, various charge and  bond ordered phases are identified
at commensurate fillings in lattice systems. 
\end{abstract} 
\maketitle

Optical traps and lattices open up a whole new direction in 
the study of strongly correlated systems in cold atomic physics.
In particular, they provide a controllable way to investigate
high spin physics 
by using atoms with  hyperfine spin multiplets.
For example,  polar and ferromagnetic condensations in spin-1 
bosonic systems (e.g. $^{23}$Na and $^{87}$Rb)
have been extensively studied 
both experimentally \cite{myatt1997,stamper-kurn1998} and 
theoretically \cite{ho1998}.
Spin nematic orders and the exotic spin liquid behavior 
are arousing much interest
\cite{demler2002, zhou2001, imambekov2003,zhou2003}
in Mott-insulating phases.
On the other hand, high spin fermionic systems also exhibit 
many properties different from those in the usual spin 1/2 
systems.
For example, Cooper pairing  shows new structures
with total spin $S\ge 2$ in the $s$-wave channel \cite{ho1999,yip1999}. 

High spin systems also provide a wonderful opportunity to investigate
the multi-particle clustering (MPC) instability and its competition
with Cooper pairing in cold atomic physics.
Taking into account the recent exciting achievement of 
fermionic superfluids by using Feshbach resonances
\cite{ho2004}, the MPC instability will  be 
one of the next focus directions.
For example, the three-body Efimov bound states in bosonic systems have
attracted wide attention \cite{braaten2004}.
In fermionic systems, the MPC instability is forbidden 
in two-component spin 1/2 systems due to Pauli's exclusion principle,
but is allowed in high spin systems with multiple components.
Knowledge learned from the MPC phase in cold atomic systems
will also shed light on
the $\alpha$-particle formation \cite{ropke1998} 
in nuclear physics, which is a 
four-fermion bound state.
Some previous theoretical works have studied the MPC phase in the 
$SU(N)$ symmetric fermionic models
by the Bethe ansatz  at 1D \cite{schlottmann1994} and a variational
method at high dimensions \cite{stepaneko1999}.
However, the stability of the MPC phase remains to be
an open problem when systems do not possess the $SU(N)$
symmetry,
and the nature of the transition between the MPC and Cooper pairing
phases has not been clarified before.

Spin 3/2 systems provide an ideal starting point to study the simplest 
MPC phase in fermionic systems, i.e., the quartetting phase, which can
be realized by using $^{132}$Cs, $^9$Be, $^{135}$Ba, $^{137}$Ba atoms.
In this article, we investigate the quartetting phase in one
dimensional (1D) spin 3/2
systems where its strong coupling nature  can be handled by 
applying the methods of bosonization and renormalization group (RG).
The investigation is greatly facilitated by the generic $SO(5)$ symmetry 
recently identified
in spin 3/2 systems \cite{wu2003a,chen2005}.
We find rich phase structures, including a gapless Luttinger
liquid  phase and two distinct spin gap phases at incommensurate fillings.
One of these spin gap phases is characterized by the quartet formation,
while the other is dominated by Cooper pairing.
The transition between them
is controlled by an Ising symmetry breaking
in one of the spin channels.
Both quartets and Cooper pairs can further undergo either the
quasi-long range ordered (QLRO) superfluidity or charge 
density wave (CDW) instabilities.
In contrast, the $SU(4)$ symmetry in the previous study
\cite{schlottmann1994} 
requires four particles to form an $SU(4)$ singlet,
thus in 1D only the quartetting phase is allowed
and Cooper paring  is excluded.
Furthermore, various charge and valence bond orders are 
identified at commensurate fillings.

We begin with phase structures at incommensurate fillings.
After linearizing the spectra around the Fermi wavevector $k_f$,
we decompose fermion operators into right and
left moving parts as $\psi_{\alpha}=\psi_{R,\alpha} e^{i k_f x}+
\psi_{L,\alpha} e^{-i k_f x} (\alpha=\pm \frac{3}{2},\pm\frac{1}{2}) $.
The right (left) moving currents are classified into 
$SO(5)$'s scalar, vector and tensor currents as
$J_{R(L)}(z) =\psi^\dagger_{R(L),\alpha}(z) \psi_{R(L),\alpha}(z)$, 
$  J^a_{R(L)}(z)=\frac{1}{2}\psi^\dagger_{R(L),\alpha}(z)\Gamma^a_{\alpha\beta}
\psi_{R(L),\beta}(z) (1\le a\le 5)$, and  $J^{ab}_{R(L)}(z)=\frac{1}{2}\psi^
\dagger_{R(L),\alpha}(z)\Gamma^{ab}_{\alpha\beta} \psi_{R(L),\beta}(z)
(1\le a<b\le 5)$, where $\Gamma^a, \Gamma^{ab}$
are the $4\times 4$ Dirac matrices defined in Ref. \cite{wu2003a}.
Classified in terms of the usual spin $SU(2)$ group,
the scalars $J_{R(L)}$ are charge currents, the 5-vectors
$J^a_{R(L)}$ are spin-nematic currents with spin $S=2$, and
the 10-tensors $J^{ab}_{R(L)}$ contain two degenerate parts
with spin $S=1,3$.

Spin 3/2 systems are characterized by two independent
$s$-wave scattering parameters $g_0$ and $g_2$ in
the total spin singlet $(S_T=0)$
and quintet $(S_T=2)$ channels, respectively.
Taking these into account,
the low energy effective Hamiltonian density reads
\bea
{\cal H}_0&=& v_f \Big\{ \frac{\pi}{4} J_R J_R +\frac{\pi}{5}
(J^a_R J^a_R +J^{ab}_R J^{ab}_R) +(R\rightarrow L)  \Big\}, \nonumber \\
{\cal H}_{int}&=&   \frac{g_c}{4} J_R J_L + g_v J^a_R J^a_L
 +g_t J^{ab}_R J^{ab}_L,
\label{ham1}
\eea
where $g_{c,v,t}$ are effective dimensionless coupling constants
with their bare values given by
\bea\label{geology}
2 g_{c}&=&  g_0+ 5 g_2,  ~ 2g_{v} = g_0-3g_2, ~
2 g_{t}=-(g_0+g_2). \ \ \
\eea
We neglect the chiral interaction terms. 
Eq. \ref{ham1} is closely related to the extensively studied
two-coupled spin 1/2 Luttinger liquids \cite{balents1996,lin1998,wu2003}.
In the latter case, Fermi wavevectors in two bands are usually different,
while they are the same for all of the four components in the former case,
thus much more low energy inter-band interactions 
are allowed.
An even larger symmetry $SU(4)$ can be obtained by fine
tuning $g_0=g_2$, i.e., $g_v=g_t$,
which means that the four spin components are equivalent.
In this case, $J^a_{R(L)}$ and $J^{ab}_{R(L)}$ form the generators of
the $SU(4)$ group.
Then Eq. \ref{ham1} reduces to the single
chain $SU(4)$ spin-orbit model \cite{azaria1999}.

\begin{figure}
\centering\epsfig{file=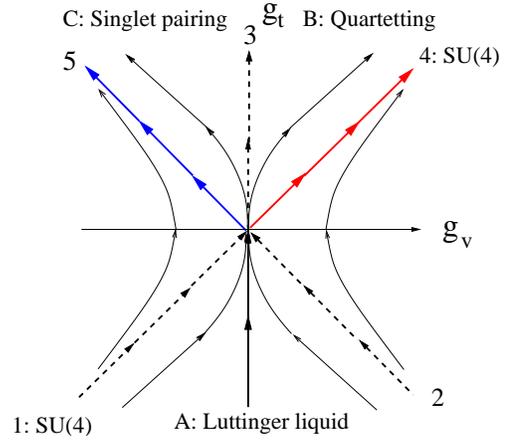,clip=1,width=0.75\linewidth,angle=0} 
\caption{ RG flows in the spin channels. 
Various phases are determined as:
A) the gapless Luttinger liquid phase;
B) the quartetting phase with QLRO superfluidity at $K_c>2$
or $2k_f$-CDW at $K_c<2$;
C) the singlet pairing phase with QLRO superfluidity at $K_c<1/2$
or $4k_f$-CDW at $K_c>1/2$.
They are controlled by the FPS of
$(0,0)$,$( +\infty, +\infty)$ (line 4), and $(-\infty, +\infty)$ (line 5) 
respectively.
Phase boundaries (line 1, 2, 3) are marked with dashed lines.
}\label{RGFig}
\end{figure}

\begin{figure}
\centering\epsfig{file=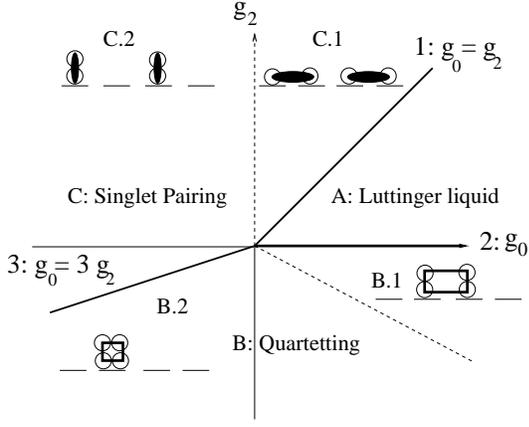,clip=1,width=0.8\linewidth,angle=0}
\caption{ Phase diagram in terms of  $g_0$ and $g_2$.
We also show the charge and bond ordered states at quarter filling
in lattice systems.
Phase A is the $SU(4)$ gapless spin liquid.
Both phase B and C split into two parts.
With $g_u\rightarrow+\infty$, B.1) quartets with both bond and
charge orders, C.1) dimerization of spin Peierls order.
With $g_u\rightarrow-\infty$, B.2) CDW phases of quartets and C.2)
CDW of singlet Cooper pairs.
Boundaries among  A, B and C are marked with solid lines,
and those between  B.1 and B.2, and C.1 and C.2
are sketched with dashed lines.
}\label{phase}
\end{figure}

The RG equations can be derived through the current 
algebra and operator product expansion  techniques \cite{senechal1999}.
The charge current satisfies the $U(1)$ Kac-Moody  algebra,
and thus $g_c$ is not renormalized at one loop level due to
the absence of Umklapp terms.
The vector and tensor currents in spin channels 
form the $SU(4)$ Kac-Moody algebra,
which gives rise to
\bea\label{RG}
\frac{ d g_v} {d  \ln (L/a) }= \frac{4}{2\pi}  g_v g_t, ~~
\frac{ d g_t}{d \ln (L/a) }= \frac{1}{2\pi} ( 3 g_t^2 + g_v^2 ),
\eea
where $L$ is the length scale and $a$ is the short distance cutoff.
The $SU(4)$ symmetry is preserved in the RG process
along the line $g_v=g_t$.
Eq. \ref{RG}  can be integrated as 
$|g_t^2-g_v^2|= c |g_v|^{3/2}~~\mbox{ ($c$: constant)}$
with the RG flows as shown in Fig \ref{RGFig}.
The parameter space $(g_v, g_t)$ is divided
into three phases as
A) the gapless Luttinger liquid,  B) and C) two different spin gap
phases with the formations of quartets and singlet pairs
respectively, which will be clarified below.
Phase A lies between $g_t=g_v<0$ (line 1) and $g_t=-g_v<0$ (line 2), 
where RG flows  evolve to the fixed point (FP)
$g_v=g_t=0$.
Phase B is bounded by line 2 and $g_v=0$ (line 3) where  RG flows
evolve to the $SU(4)$ marginally relevant FP of
$g_v=g_s\rightarrow +\infty$ (line 4).
Phase C is symmetric to B under the reflection $g_v\rightarrow -g_v$.
RG flows evolve to the line 5 with $-g_v=g_t\rightarrow +\infty$. 
The boundary between B and C, i.e., line 3,  is controlled by the unstable
FP of $g_v=0, g_t\rightarrow +\infty$.
From the relations in Eq. \ref{geology},
we replot the phase diagram 
in terms of the $s$-wave scattering lengths
$g_0$ and $g_2$ as shown in Fig. \ref{phase}.
The boundary between phase A and C, i.e., $g_0=g_2>0$,
is exact from  the $SU(4)$ symmetry 
regardless of the one-loop approximation.
On the contrary, boundaries between A and B, B and C
will be modified in higher order perturbations.

To clarify the nature of each phase, we employ the Abelian bosonization.
The bosonization identity reads
$\psi_{R(L)\alpha} (x)= \eta_\alpha/  \sqrt{2\pi a} \exp \{\pm i \sqrt{\pi} 
(\phi_\alpha(x)\pm \theta_\alpha(x))\}
(\alpha=\pm\frac{3}{2},\pm\frac{1}{2})$  where $\eta s$ are the Klein factors.
Boson fields $\phi_\alpha$ and their dual fields $\theta_\alpha$ 
are conveniently reorganized into $\phi_c(\theta_{c})$ in 
the charge channel, and $\phi_v(\theta_v), \phi_{t1}(\theta_{t1}),
\phi_{t2}(\theta_{t2})$ in the spin channels via
$
\phi_{c,v}=(\phi_{\frac{3}{2}}
\pm\phi_{\frac{1}{2}}\pm\phi_{-\frac{1}{2}} +\phi_{-\frac{3}{2}})/2,
\phi_{t1,t2}=(\phi_{\frac{3}{2}}\mp\phi_{\frac{1}{2}}
\pm\phi_{-\frac{1}{2}}-\phi_{-\frac{3}{2}})/2.
$
Similar expressions also hold for $\theta$s.
The quadratic part of the Hamiltonian density
is standard  $(\nu=c,v,t_1,t_2)$
\bea 
{\cal H}_0&=&\frac{v_\nu}{2}\sum_\nu \Big\{ K_\nu (\partial_x \theta_\nu)^2 + 
\frac{1}{K_\nu} (\partial_x \phi_\nu)^2 \Big \},
\eea
with Luttinger parameters $K_\nu$ and velocities $v_\nu$ in each channel.
The non-quadratic terms are summarized as 
\bea\label{CosTerms}
{\cal H}_{int}&=&-  \frac{1}{2(\pi a)^2} \big \{\cos\sqrt {4\pi} \phi_{t1}+
\cos\sqrt {4\pi} \phi_{t2} \big\} \nonumber \\
&\times& 
\big\{ (g_t+g_v) \cos\sqrt{4\pi}\phi_v  +(g_t-g_v) \cos\sqrt{4\pi}\theta_v
\big\} \nonumber \\
&-& \frac {g_t} { 2(\pi a)^2} 
\cos\sqrt {4\pi} \phi_{t1} \cos\sqrt {4\pi} \phi_{t2},
\eea
with the convention of  Klein factors as $\eta_{\frac{3}{2}} 
\eta_{\frac{1}{2}} \eta_{\frac{-1}{2}} \eta_{\frac{-3}{2}}=1$.

Various order parameters are needed to characterize phase structures,
including the  $2k_f$-CDW operator $N$,  the spin
density wave (SDW) operators  in the SO(5) vector channel $N^a$ 
and their tensor channel version $N^{ab}$, the singlet pairing operator
$\eta$ and its quintet counterpart  $\chi^a$, 
the $4k_f$-CDW operator  $O_{4k_f,cdw}$, 
as well as the quartetting operator $O_{qt}$.
They are defined as
\bea\label{eq:order}
&& N=\psi^\dagger_{R\alpha}  \psi_{L\alpha},  
N^a= \psi^\dagger_{R\alpha} \frac{\Gamma^a_{\alpha\beta} }{2}
\psi_{L\beta}, N^{ab}= \psi^\dagger_{R\alpha}
\frac{\Gamma^{ab}_{\alpha\beta}}{2} \psi_{L\beta}; \nonumber\\
&&\eta^\dagger=\psi^\dagger_{R,\alpha} R_{\alpha\beta}\psi^\dagger_{L,\beta},
\ \ \
\chi^{a,\dagger}=i
\psi^\dagger_{R,\alpha}(R\Gamma^a)_{\alpha\beta}\psi^\dagger_{L,\beta};
\nonumber \\
&&O_{4k_f,cdw}=\psi^\dagger_{R\alpha} \psi^\dagger_{R\beta} 
\psi_{L\beta} \psi_{L\alpha}, \nonumber \\
&&O_{quar}= (\epsilon_{\alpha\beta\gamma\delta}/4)
\psi^\dagger_{R\alpha} \psi^\dagger_{R\beta} 
\psi^\dagger_{L\gamma} \psi^\dagger_{L\delta},
\label{eq:orders}
\eea
where $R=\Gamma_1 \Gamma_3$ \cite{wu2003} is the charge conjugation matrix
and $\epsilon_{\alpha\beta\gamma\delta}$ is the rank-4 Levi-Civita symbol.

Now we are ready to discuss the nature in each phase. 
In the Luttinger liquid phase A, all the cosine terms are marginally 
irrelevant.
As shown in Fig. \ref{phase}, the Luttinger phase exists 
in the repulsive region
with $g_0\ge g_2\ge 0$, thus  $K_c<1$.
As a result, only $2k_f$-CDW and SDW  susceptibilities diverge
with scaling dimensions $\Delta_{N}=\Delta_{N^a}=\Delta_{N^{ab}}=(K_c+3)/4$.
Taking into account the chiral interaction terms,
two different spin velocities exist in the vector and tensor
channels  as exhibited in the one-particle correlation functions 
$\avg{|\psi_{R(L),\alpha}(x,\tau)\psi^\dagger_{R(L),\alpha}(0,0)|}$ 
\bea 
\frac{1}{(v_c \tau \mp i x)^{\frac{1}{4}} }
\frac{1}{(v_v \tau \mp i x)^{\frac{1}{4}} }
\frac{1}{(v_t \tau \mp i x)^{\frac{1}{2}} } 
\Big(\frac{a}{|v_c \tau -i x|}\Big)^{\gamma_c} ~~ 
\eea
with  $\gamma_c=\frac{1}{8} (K_c+ 1/K_c-2)$.

Phase B is characterized by spin gaps in $\phi_v$, $\phi_{t1}$
and $\phi_{t2}$ channels.
Its $SU(4)$ FP of $g_v=g_t\rightarrow+\infty$ means
the formation of the $SU(4)$ invariant quartets as follows.
In the ground state, cosine operators acquire non-vanishing expectation
values as $\avg{\cos\sqrt{4\pi}\phi_v}
=\avg{\cos\sqrt{4\pi}\phi_{t1}}=\avg{\cos\sqrt{4\pi}\phi_{t2}}$,
whose classical values are just $1$ or $-1$ as  related by a $Z_2$
symmetry \cite{fjaerestad2002}.
We fix the gauge by choosing  $\avg{\phi_v}=\avg{\phi_{t1}}
=\avg{\phi_{t2}}=0$.
By checking scaling dimensions of various order parameters
in Eq. \ref{eq:orders},
we conclude that the competing instabilities are the QLRO superfluidity 
$O_{quar}$ and CDW of quartets.
Because the average distance between two quartets is $d=\pi/k_f$,
this  CDW is of $2k_f$ type.
Their expressions are reduced to 
$O_{quar}\propto e^{2i\sqrt{\pi} \theta_c}$ 
and
$N\propto e^{i\sqrt \pi \phi_c}$
respectively,
with scaling dimensions
$\Delta_{quar}= 1/K_c$ and
$\Delta_{N}=\frac{1}{4} K_c$.
The leading instability is 
$O_{quar}$ at $K_c>2$, and  $N$ at $K_c<2$ respectively.
We extend the quartets formation regime from
the previous Bethe-ansatz results along 
$SU(4)$ line 4  \cite{schlottmann1994}
to the entire phase B.
On the other hand,  correlations of pairing operators
$\eta^\dagger$ and 
$\chi^{a\dagger}$ decay exponentially.

Phase C is controlled by the FP of $-g_v=g_t\rightarrow+\infty$
which is characterized by the singlet Cooper pairing, and
possesses another $SU(4)$ symmetry denoted as $SU^\prime(4)$.
Its right (left) generators $J^\prime_{R(L)}$ belong to the 
$SU(4)$ fundamental (anti-fundamental) representations 
defined as
\bea
J^{\prime,ab}_{R}=J^{ab}_{R}, J^{\prime,a}_{R}= J^{a}_{R}; \ \ \
J^{\prime,ab}_{L}=J^{ab}_{L}, J^{\prime,a}_{L}=-J^{a}_{L}.
\eea
It is different from the $SU(4)$ symmetry in phase B where both right and
left generators are in the fundamental representation.
The singlet Cooper pair is invariant under this $SU^\prime(4)$ symmetry
but not the $SU(4)$ symmetry in phase B.
We  choose the pinned bosonic fields to be
$\avg{\phi_{t1}}=\avg{\phi_{t2}}=\avg{\theta_v}=0$,
where the dual field $\theta_v$ instead of $\phi_v$ is pinned
as in Phase B.
Again, the superfluidity and CDW orders of pairs compete.
Because the average distance between adjacent pairs is $d=\pi/(2k_f)$,
this CDW is of the $4k_f$ type.
These two orders are reduced to
$\eta^\dagger\propto e^{-i\sqrt \pi \theta_c}$ and
$O_{4k_f,cdw}\propto e^{i\sqrt{4\pi} \phi_c}$, 
with scaling dimensions as $\Delta_{\eta}= 1/(4 K_c)$
and $\Delta_{4k_f, cdw}=K_c$.
The leading instability is $\eta^\dagger$ at $K_c>1/2$, and
$O_{4k_f,cdw}$ at $K_c<1/2$ respectively. 
Remarkably, due to the presence of the spin gap, the singlet pairing
instability dominates even in the purely repulsive region at $g_0>g_2>0$
when  $K_c>1/2$ as shown in Fig. \ref{phase}.
This is similar to the situation in high T$_c$ superconductivity.

The transition between quartetting and pairing  phases is controlled 
by an Ising duality in the $\theta_v (\phi_v)$ channel.
Near the phase boundary, $\phi_{t1,2}$
are pined, thus the residual interaction becomes
\bea
H_{res}&=&- \frac{\lambda} {2(\pi a)^2} \Big\{
(g_t+g_v) \cos\sqrt{4\pi}\phi_v + (g_t-g_v)\nonumber\\
&\times&\cos\sqrt{4\pi}\theta_v\Big\},
\label{eq:res}
\eea
where $\lambda=- \frac{1}{\pi a} (\avg {\cos\sqrt{4\pi}\phi_{t1}}
+\avg{\cos{\sqrt4\pi}\phi_{t2}} )$.
The singlet pairing operator $\eta^\dagger 
\propto \psi^\dagger_{\frac{3}{2}}\psi^\dagger_{\frac{-3}{2}}
-\psi^\dagger_{\frac{1}{2}}\psi^\dagger_{\frac{-1}{2}}
\propto e^{-i\sqrt \pi\theta_c} (e^{-i\sqrt\pi\theta_v}
+e^{i\sqrt\pi\theta_v})$, thus $\sqrt{4\pi}\theta_v$ is the relative phase 
between the pairs of $\Delta^\dagger_1=\psi^\dagger_{\frac{3}{2}}
\psi^\dagger_{-\frac{3}{2}}$ and  $\Delta^\dagger_2=
\psi^\dagger_{\frac{1}{2}}\psi^\dagger_{-\frac{1}{2}}$,
and $\phi_v$ is the vortex field dual to $\theta_v$.
The transition can be viewed as the phase locking problem in 
a two-component superfluids.
In the pairing phase, $\theta_v$ can be pinned 
at either $0$ or $\sqrt\pi$ by the internal Josephson
term $\cos\sqrt{4\pi}\theta_v$.
This symmetry is $Z_2$, and thus the pairing phase with
pinned $\theta_v$ is in the Ising ordered phase.
In contrast, the quartetting phase is the Ising disordered phase
where the dual field $\phi_v$ is locked instead. 
The Ising nature of the transition is also clear by representing
Eq. \ref{eq:res} in terms of two Majorana fermions 
\cite{schulz1996} as
\bea
H_{res}&=& \sum_{a=1}^2 (\xi^a_R i\partial_x \xi^a_R-
\xi^a_L i\partial_x \xi^a_L)+
i \lambda (g_t \xi^1_R \xi^1_L +g_v \xi^2_R \xi^2_L),
\nonumber
\eea
$\xi^1$ is always off-critical as $g_t\rightarrow +\infty$, 
while $g_v$ is the mass of $\xi^2$ which changes sign across the boundary.
Thus the quartetting and pairing phases are Ising disordered
and ordered  phases for $\xi^2$, respectively.

Next we consider the charge and valence bond ordered state
as depicted in Fig. \ref{phase} at quarter-filling, 
i.e, one particle per site.
The $8k_f$ Umklapp term appears, and decouples with spin channels,
which can be bosonized as
\bea\label{um8kf}
{\cal H}^\prime_{um}&=& \frac{g_u}{2(\pi a)^2} \cos (\sqrt {16 \pi}
\phi_{c}-8 k_f x ),
\eea 
where  $g_u$  is at the order of $O(g^3_0,g^3_2)$ at the bare level.
At $K_c<1/2$, this term is relevant and opens the charge gap.
The real parts of  $N$ and $O_{4k_f,cdw}$ 
describe the usual CDW orders, while their imaginary parts 
mean the bond orders, i.e., the $2k_f$ and $4k_f$ spin Peierls orders.
Phase A remains gapless in spin channels as described by the 
$SU(4)$ Heisenberg model.
The quartetting phase B splits into two parts B.1 and B.2 with 
$g_u \rightarrow +\infty (-\infty)$ respectively.
In B.1, $\Im O_{4k_f,cdw}$, $\Re N$ and 
$\Im  N$ are long range ordered.
The $SU(4)$ singlet quartets exhibit both charge and
spin Peierls orders.
Instead, the 2$k_f$-CDW of quartets becomes long range ordered
in B.2.
Similarly, the singlet pairing phase C splits into C.1 and C.2
as $g_u \rightarrow +\infty (-\infty)$ respectively.
In C.1, $\Im O_{4k_f,cdw}$, i.e., the spin 
Peierls order is stabilized.
Instead, the CDW of singlet pairs becomes long range ordered
in C.2. 
The boundaries between phase B.1 and B.2, phase C.1 and C.2 are
determined by the bare value of $g_{u0}=0$ as sketched
in Fig. \ref{phase}.
However, due to the non-universal relation between $g_{u0}$ and
$g_{0,2}$, the exact boundaries are hard to determine.

The model discussed above can be realized by loading
spin 3/2 fermions into one-dimensional optical tubes.
The quartetting phase can be probed by the radio-frequency 
spectroscopy to measure
the excitation gaps of the successive quartet breaking process as
$4\rightarrow 1+3 \rightarrow 1+1+2 \rightarrow 1+1+1+1$.
In contrast, in the pairing phase, only one pairing breaking
process of $2\rightarrow 1+1$ exits as
measured experimentally \cite{bartenstein2004}.
The charge and bond ordered states break 
translational symmetry.
The periodicity of phases $C1$ and $C2$
is two lattice constants,
while that of $B1$ and $B2$ is doubled.
These can be  detected by using the 
elastic Bragg spectroscopy \cite{oosten2004}.

In summary, we have investigated the global phase diagram 
in 1D spin 3/2 systems, including the gapless Luttinger liquid,
the spin gapped quartetting and Copper pairing phases.
The competition between the quartetting and pairing phases
is found to be controlled by an Ising duality.
Both quartets and pairs can either undergo QLRO superfluidity
or CDW instabilities, depending on the value of Luttinger 
parameter $K_c$.
The QLRO singlet pair superfluidity can survive in the purely 
repulsive interaction regime.
The Mott-insulating phases at commensurate fillings exhibit
various charge and valence bond orders.

{\it Note added}\ \ \
After the paper was submitted for publication, we became aware of 
the work by Lecheminant, Azaria, and
Boulat \cite{azaria2004},
where the two spin gap phases, and Mott phases at quarter filling
are obtained independently.

C. W. would like to express his gratitude to  E. Demler, A. J. Leggett
and S. C. Zhang for their introduction on the quartetting instability.
C. W. also thanks S. Capponi, E. Fradkin, J. P. Hu, and 
P. Lecheminant for helpful 
discussions and R. K. Elliott for his improving
the manuscript.
This work is supported by the NSF under grant numbers DMR-9814289, and
the US Department of Energy, Office of Basic Energy Sciences under 
contract DE-AC03-76SF00515.  C.W. is also supported by the
Stanford SGF.

\end{document}